\documentclass[equations,11pt]{article}
\usepackage{amsfonts}
\usepackage{a4,tikz}
\usepackage{geometry}
\usepackage{color} 
\usepackage{subfigure}
\makeatletter
\newcommand\ackname{Acknowledgements}
\if@titlepage
  \newenvironment{acknowledgements}{%
      \titlepage
      \null\vfil
      \@beginparpenalty\@lowpenalty
      \begin{center}%
        \bfseries \ackname
        \@endparpenalty\@M1
      \end{center}}%
     {\par\vfil\null\endtitlepage}
\else
  \newenvironment{acknowledgements}{%
      \if@twocolumn
        \section*{\abstractname}%
      \else
        \small
        \begin{center}%
          {\bfseries \ackname\vspace{-.5em}\vspace{\z@}}%
        \end{center}%
        \quotation
      \fi}
      {\if@twocolumn\else\endquotation\fi}
\fi
\makeatother
\usepackage{amsmath}
\usepackage{amssymb}
\usepackage{amsthm}
\usepackage{mathrsfs}
\usepackage{eucal}
 \usepackage[usenames,dvipsnames]{pstricks}
 \usepackage{epsfig}
 \usepackage{pst-grad} % For gradients
 \usepackage{pst-plot} % For axes
\renewcommand{\theequation}{\arabic{equation}}
\usepackage{dsfont}
\usepackage{amssymb,amsmath}
\usepackage{color}
\usepackage{accents}
\usepackage{texdraw}
\usepackage{eucal}
\usepackage{epic,epsfig}
\usepackage{graphicx}

\theoremstyle{definition}

\numberwithin{equation}{section}
\DeclareMathAccent{\wtilde}{\mathord}{largesymbols}{"65}
\DeclareMathAccent{\what}{\mathord}{largesymbols}{"62}

\def\m@th{\mathsurround=0pt}
\mathchardef\bracell="0365
\def\upbrall{$\m@th\bracell$}
\def\undertilde#1{\mathop{\vtop{\ialign{##\crcr
    $\hfil\displaystyle{#1}\hfil$\crcr
     \noalign
     {\kern1.5pt\nointerlineskip}
     \upbrall\crcr\noalign{\kern1pt
   }}}}\limits}
\def\m@th{\mathsurround=0pt}
\mathchardef\bracell="0365
\def\upbrall{$\m@th\bracell$}
\def\underhat#1{\mathop{\vtop{\ialign{##\crcr
    $\hfil\displaystyle{#1}\hfil$\crcr
     \noalign
     {\kern1.5pt\nointerlineskip}
     \upbrall\crcr\noalign{\kern1pt
   }}}}\limits}
\usepackage{pgf}
\usepackage{tikz}
\usepackage{verbatim}
\usepackage[utf8]{inputenc}
\usetikzlibrary{arrows,automata}
\usetikzlibrary{positioning}

\def\theequation{\arabic{section}.\arabic{equation}}

\newcommand{\wh}{\widehat}
\newcommand{\wt}{\widetilde}

\def\hypotilde#1#2{\vrule depth #1 pt width 0pt{\smash{{\mathop{#2}
\limits_{\displaystyle\widetilde{}}}}}}
\def\hypohat#1#2{\vrule depth #1 pt width 0pt{\smash{{\mathop{#2}
\limits_{\displaystyle\widehat{}}}}}}

\newcommand{\bblu}{\begin{color}{blue}}
\newcommand{\bred}{\begin{color}{red}}
\newcommand{\ecl}{\end{color}}

\newcommand{\bLam}{\boldsymbol{\Lambda}}
\newcommand{\be}{\begin{equation}}
\newcommand{\ee}{\end{equation}}
\newcommand{\bea}{\begin{eqnarray}}
\newcommand{\eea}{\end{eqnarray}}
\newcommand{\bse}{\begin{subequations}}
\newcommand{\ese}{\end{subequations}}
\newcommand{\nn}{\nonumber}

\begin{document}

\def\theequation{\arabic{section}.\arabic{equation}}

\newtheorem{thm}{Theorem}[section]
\newtheorem{lem}{Lemma}[section]
\newtheorem{defn}{Definition}[section]
\newtheorem{ex}{Example}[section]
\newtheorem{rem}{}
\newtheorem{criteria}{Criteria}[section]
\newcommand{\ra}{\rangle}
\newcommand{\la}{\langle}
%\newcommand{\be}{\begin{equation}}
%\newcommand{\ee}{\end{equation}}
%\newcommand{\bea}{\begin{eqnarray}}
%\newcommand{\eea}{\end{eqnarray}}
%\newcommand{\bse}{\begin{subequations}}
%\newcommand{\ese}{\end{subequations}}
%\newcommand{\nn}{\nonumber}
%\input{tcilatex}

%%%%%%%%%%%%%%%%%%%%%%%%%%%%%%%%%%%%%%%%%%%%%%%%%%%%%%%%%%%%%%%%%%%%%%%%%%%%
\title{\textbf{One-parameter discrete-time Calogero-Moser system}}
\author{\\\\Umpon Jairuk$^\dagger$ and Sikarin Yoo-Kong$^{*} $ \\
\small $^\dagger $\emph{Division of Physics, Faculty of Science and Technology,}\\ 
\small \emph{Rajamangala University of Technology Thanyaburi, Rangsit-Nakornnayok Road,}\\
\small \emph{Pathumthani, Thailand 12110.}\\
\small $^* $ \emph{The Institute for Fundamental Study(IF), Naresuan University(NU), }\\
\small \emph{99 Moo 9, Tha Pho, Mueang Phitsanulok, Phitsanulok, Thailand, 65000}\\
}
\maketitle
%%%%%%%%%%%%%%%%%%%%%%%%%%%%%%%%%%%%%%%%%%%%%%%%%%%%%%%
%%%%%%%%%%%%%%%%%%%%%%%%%%%%%%%%%%%%%%%%%%%%%%%%%%%%%%%

%%%%%%%%%%%%%%%%%%%%%%%%%%%%%%%%%%%%%%%%%%%%%%%%%%%%%%%

\abstract
We present a new type of integrable one-dimensional many-body systems called a one-parameter Calogero-Moser (CM) system. In the discrete level, the Lax pairs with a parameter are introduced and, of course, the discrete-time equations of motion are obtained as well as their corresponding discrete-time Lagrangian. The integrability feature of this new system can be captured through the discrete Lagrangian closure relation by employing a connection with the temporal Lax matrices of the discrete-time Ruijsenaars-Schneider (RS) system, exact solution, and the existence of the classical r-matrix. 
%The continuum limit is considered and the one-parameter CM system is obtained in terms of the equations of motion and the Lagrangian. 
Under the appropriate limit on the parameter, which in this case is approaching zero, the standard CM system is retrieved both discrete-time and continuous-time.
%It is obtained by adding one parameter in Lax matrices. compatibility between two discrete flows leads to a discrete closure relation for the corresponding Lagrangians moreover we find the connection between Lagrangian and M-matrix of the discrete-time Ruijsenaars-Schneider(RS) system. The continuous-time one-parameter CM system is obtained through the continuum limits. We notice that when the parameter is reduced to 0 in equations of motion, we receive equations of motion of rational CM system in both discrete and continuous cases. 
%Starting with one parameter, we introduce a new discrete integrability system which is a map of the Calogero-Moser system. In this framework, the limit of the parameter to 0 leads to discrete rational Calogero-Moser system.
%We study teo guarantee effectiveness of performing on discrete-time to continuous time in 1-from structure. There are two steps\cite{Sikarin1, Sikarin2, Umpon} for obtaining Lagrangian hierarchies. To confirm this method, we consider Hyperbolic Calogerlo-Moser systems consisting of Hyperbolic Calogero-Moser(HCM), Hyperbolic Ruijsenaaars-Schneider(HRS), and Hyperbolic Goldfish(HGF) system.  
%%%%%%%%%%%%%%%%%%%%%%%%%%%%%%%%%%%%%%%%%%%%%%%%%%%%%%%%
%%%%%%%%%%%%%%%%%%%%%%%%%%%%%%%%%%%%%%%%%%%%%%%%%%%%%%%%

%%%%%%%%%%%%%%%%%%%%%%%%%%%%%%%%%%%%%%%%%%%%%%%%%%%%%%%%
\section{Introduction}\label{intro}
\setcounter{equation}{0}
The Calogero-Moser (CM) system is a mathematical model that describes the motion of one-dimensional system of particles interacting through long-range forces \cite{Cal, Mos}. The CM system is, of course, an integrable system which exhibits rich symmetries and possesses a sufficient number of conserved quantities, according to Liouville's integrability notion, to construct the exact solutions. Let us give the equations motion of the CM system for the simplest type of interaction, known as the rational case,
\begin{equation}
\ddot x_i=\sum_{j=1}^N\frac{1}{(x_i-x_j)^3}\;,\;\;\;i=1,2,3,...,N\;,
\end{equation}
where $x_i$ is a position of the $i^{th}$ particle.
\\
\\
The Ruijsenaars-Schneider (RS) system is another integrable one-dimensional system of particles with a long-range interaction\cite{Raj, Sch}. In the simplest interaction, namely the rational case, the equations of motion are given by
\begin{equation}
    \ddot x_i+\sum_{j=1}^N\dot x_i\dot x_j\left(\frac{1}{x_i-x_j+\lambda}+\frac{1}{x_i-x_j-\lambda}-\frac{2}{x_i-x_j}\right)\;,\;\;\;i=1,2,3,...,N;,
\end{equation}
where $\lambda$ is a parameter. Under the limit: $\lambda \to 0$, the CM system is recovered. Then the RS system can be treated as a ``one-parameter generalisation" of the CM system.
\\
In 1994, the time-discretised version of the CM system was introduced by Nijhoff and Pang \cite{Nij-Pan}. In the rational case, the discrete-time equations of motion is given by
\begin{equation}
\sum\limits_{k = 1}^N \left(\frac{1}{x_i-\wt{x}_k}+\frac{1}{x_i-{\hypotilde 0 x}_k} \right) - \sum\limits_{k = 1 \atop k \neq i }^N \frac{2}{x_i-x_k} = 0, 
\end{equation}
where $\wt x_i=x_i(n+1)$ is a forward shift and ${{\hypotilde 0 x}_i}=x_i(n-1)$ is a backward shift. The integrability of the system can be captured in the same sense with the continuous system in terms of the classical r-matrix, the existence of the exact solution, and the existence of a set of sufficient invariants.
\\
\\
Soon after, the time-discretised version of the RS system was introduced \cite{Nij-R-K}. In the rational case, the discrete-time equations of motion are given by
\begin{equation}
    \prod_{j=1\atop jk \neq i}^N\frac{x_i-x_j+\lambda}{x_i-x_j-\lambda}=\prod_{j=1}^N\frac{(x_j-\wt x_j)(x_i-{{\hypotilde 0 x}_j}+\lambda)}{(x_j-{{\hypotilde 0 x}_j})(x_i-\wt x_j+\lambda)}\;.
\end{equation}
Again, under the limit: $\lambda \to 0$, the discrete-time CM system is recovered. Of course, the discrete-time RS system can also be treated as the ``one-parameter generalisation" of the discrete-time CM system.
\\
\\
Recently, a new hallmark for integrability was promoted known as the multi-dimensional consistency. On the level of the discrete-time equations of motion, the multi-dimensional consistency can be inferred as the consistency around the cube \cite{NijWal, Nij}. On the level of the Hamiltonians, the feature can be captured through the Hamiltonian commuting flows as a direct consequence of the involution in Liouville's integrability \cite{Babelon}. Alternatively, on the level of Lagrangians, the multi-dimensional consistency can be expressed through the Lagrangian closure relation as a direct result in the variation of the action with respect to independent variables. Since the closure relation for Lagrangian 1-form will play a major role in this paper as an integrability criterion, then we shall spend a bit more space to derive the relation.
Now let $\boldsymbol{n}$ be a vector in the lattice and let $\boldsymbol{e}_i$ be a unit vector in the $i^{th}$ direction. Then an elementary shift in the $i^{th}$ direction on the lattice is defined as $\boldsymbol{n} \to\boldsymbol{n}+\boldsymbol{e}_i$. Therefore, the discrete-time Lagrangians can be expressed in the form
\begin{equation}
    \mathcal{L}_i(\boldsymbol{n})=\mathcal{L}_i(\boldsymbol{x}(\boldsymbol{n}),\boldsymbol{x}(\boldsymbol{n}+\boldsymbol{e}_i))\;,
\end{equation}
where $\boldsymbol{x}=\{x_1,x_2,...,x_N\}$. The discrete-time action is defined as
\begin{equation}
    S=S[\boldsymbol{x}(\boldsymbol{n}):\Gamma]=\sum_{\boldsymbol{n}\in \Gamma}\mathcal L_i(\boldsymbol{x}(\boldsymbol{n}),\boldsymbol{x}(\boldsymbol{n}+\boldsymbol{e}_i))\;,
\end{equation}
where $\Gamma$ is a arbitrary discrete curve, see figure \ref{F0}. Next, we shall consider another discrete curve $\Gamma'$ sharing the same endpoints with the discrete curve $\Gamma$ and the action is given by
\begin{equation}
    S'=S[\boldsymbol{x}(\boldsymbol{n}):\Gamma']=\sum_{\boldsymbol{n}\in \Gamma'}\mathcal L_i(\boldsymbol{x}(\boldsymbol{n}),\boldsymbol{x}(\boldsymbol{n}+\boldsymbol{e}_i))\;.
\end{equation}
Of course, this can be viewed as the variation of independent variables $\boldsymbol{n}\to \boldsymbol{n}+\Delta\boldsymbol{n}$ of the action
\begin{eqnarray}
    S'&=&S-\mathcal L_i(\boldsymbol{x}(\boldsymbol{n}+\boldsymbol{e}_j),\boldsymbol{x}(\boldsymbol{n}+\boldsymbol{e}_i+\boldsymbol{e}_j))+\mathcal L_i(\boldsymbol{x}(\boldsymbol{n}),\boldsymbol{x}(\boldsymbol{n}+\boldsymbol{e}_i))\nn\\
    &&+\mathcal L_j(\boldsymbol{x}(\boldsymbol{n}+\boldsymbol{e}_i),\boldsymbol{x}(\boldsymbol{n}+\boldsymbol{e}_j+\boldsymbol{e}_i))-\mathcal L_j(\boldsymbol{x}(\boldsymbol{n}),\boldsymbol{x}(\boldsymbol{n}+\boldsymbol{e}_j))\;.
\end{eqnarray}
The least action principle requires $\delta S=S'-S=0$ resulting in
\begin{eqnarray}
    0&=&\mathcal L_i(\boldsymbol{x}(\boldsymbol{n}+\boldsymbol{e}_j),\boldsymbol{x}(\boldsymbol{n}+\boldsymbol{e}_i+\boldsymbol{e}_j))-\mathcal L_i(\boldsymbol{x}(\boldsymbol{n}),\boldsymbol{x}(\boldsymbol{n}+\boldsymbol{e}_i))\nn\\
    &&-\mathcal L_j(\boldsymbol{x}(\boldsymbol{n}+\boldsymbol{e}_i),\boldsymbol{x}(\boldsymbol{n}+\boldsymbol{e}_j+\boldsymbol{e}_i))+\mathcal L_j(\boldsymbol{x}(\boldsymbol{n}),\boldsymbol{x}(\boldsymbol{n}+\boldsymbol{e}_j))\;,\label{CR1}
\end{eqnarray}
which is the closure relation for the discrete-time Lagrangian 1-form. Equivalently, for two-dimensional lattice, see figure \ref{F1}, \eqref{CR1} can be re-expressed in the form
\begin{equation}\label{OCM78}
 \widehat{\mathcal{L}(\boldsymbol{x},\widetilde{\boldsymbol{x}})} - \mathcal{L}(\boldsymbol{x},\widetilde{\boldsymbol{x}}) -
 \widetilde{\mathcal{L}(\boldsymbol{x},\widehat{\boldsymbol{x}})} + \mathcal{L}(\boldsymbol{x},\widehat{\boldsymbol{x}}) = 0\;.
\end{equation}
\begin{figure}[h] 
\begin{center}
\begin{tikzpicture}[scale=0.45]
 \draw[->] (0,0) -- (17,0) node[anchor=west] {$n_{i}$};
 \draw[->] (0,0) -- (6.5,12) node[anchor=south] {$n_{j}$};
 \fill (4,2) circle (0.15);
 \draw[dashed] (4,2) --(4.8,3.5)--(6,3.5)--(7.3,6)--(9.5,6)--(11.5,10)--(15.2,10) ;
 \fill (15.2,10) circle (0.15);
 \draw (6,7)node[anchor=north west] {$\Gamma$};
\draw (4,2) --(7,2)--(7.7,3.4)--(10,3.4)--(11,5.5)--(13,5.5)--(15.2,10) ;
\draw (13.5,7)node[anchor=north west] {$\Gamma^{\prime}$};
\end{tikzpicture}
\end{center}
\caption{Arbitrary discrete curves on the space of independent variables.}\label{F0}
\end{figure}
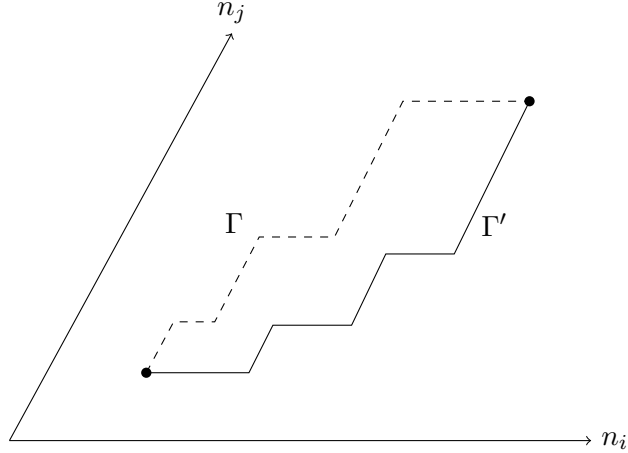
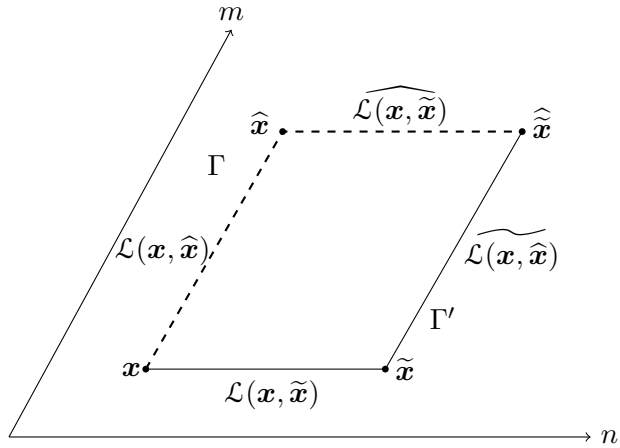
\begin{figure}[h]
\begin{center}
\begin{tikzpicture}[scale=0.45]
\draw[->] (0,0) -- (17,0) node[anchor=west] {$n$};
\draw[->] (0,0) -- (6.5,12) node[anchor=south] {$m$};
 \draw[thick,dashed,black] (4,2)--(8,9)--(15,9);
 %%%%%%%%%%%%%%%%%%%%%%%%%%%%%%%%%%%%%%%%%%%%%%%%%%%%%%%
 \fill (4,2) circle (0.1);
 \fill (8,9) circle (0.1);
 \fill (11,2) circle (0.1);
 \fill (15,9) circle (0.1);
  %%%%%%%%%%%%%%%%%%%%%%%%%%%%%%%%%%%%%%%%%%%%%%%%%%%%%%%%
 \draw (3,2) node[anchor=west] {$\boldsymbol x$};
\draw (11,2) node[anchor=west] {$\wt{\boldsymbol x}$};
 \draw (6.75,9.25) node[anchor=west] {$\wh{\boldsymbol x}$};
 \draw (15,9.25) node[anchor=west] {$\wh{\wt{\boldsymbol x}}$};
 \draw (5.5,8)node[anchor=west] {$\Gamma$};
 \draw (12,3.5)node[anchor=west] {$\Gamma^{\prime}$};
 \draw[black] (4,2) --(11,2)  -- (15,9);
 %%%%%%%%%%%%%%%%%%%%%%%%%%%%%%%%%%%%%%%%%\%%%%%%%%%%%%
\draw (9.75,9.75) node[anchor=west] {$  \widehat{\mathcal{L}(\boldsymbol{x},\widetilde{\boldsymbol{x}})}$};
\draw (2.8,5.5) node[anchor=west] {$\mathcal{L}(\boldsymbol{x},\widehat{\boldsymbol{x}})$};
\draw (6,1.25) node[anchor=west] {$\mathcal{L}(\boldsymbol{x},\widetilde{\boldsymbol{x}})$};
\draw (13.00,5.5) node[anchor=west] {$\widetilde{\mathcal{L}(\boldsymbol{x},\widehat{\boldsymbol{x}})}$};
\end{tikzpicture}
\end{center}
\caption{The local variation of the discrete curve on the space of two independent variables.}\label{F1}
\end{figure}
\\
\\
In this work, we propose a new type of one-parameter CM system, apart from the RS system and study its integrability through the existence of the exact solution, classical r-matrix and the closure relation. Therefore, the structure of the paper is in the following. In section 2, the two compatible  one-parameter discrete-time CM systems will be obtained from the Lax equations. In section 3, the discrete-time Lagrangians are also established and the closure relation is directly obtained via the connection between the RS temporal Lax matrices and the Lagrangian. In section 4, the classical r-matrix for the one-parameter discrete-time CM system is considered. In section 5, the construction of the exact solution is carefully derived. In section 6, the continuum limit will be performed on the one-parameter discrete-time CM system resulting in the one-parameter continuous-time CM system. The final section is a summary and possible further investigations. 
%%%%%%%%%%%%%%%%%%%%%%%%%%%%%%%%%%%%%%%%%%%%%%%%%%%%%
\section{One-parameter discrete-time CM system}\label{DEM}
\setcounter{equation}{0} 
In this section, we will construct the discrete-time CM system with a parameter $\lambda$. First, we introduce the spatial Lax matrix: $\boldsymbol L_{\lambda}$ with two temporal matrices: $\boldsymbol M$ and $\boldsymbol N$ as follows
\begin{subequations}\label{OCM1} 
\begin{eqnarray}
\boldsymbol L_{\lambda}&=&\sum_{i,j=1}^N\frac{1}{x_i-x_j+\lambda}E_{ij}\;,\label{OCM111}\\
\boldsymbol M&=&\sum_{i,j=1}^N\frac{1}{\wt{x}_i-x_j}E_{ij}\;,\label{OCM121}\\
\boldsymbol N&=&\sum_{i,j=1}^N\frac{1}{\wh{x}_i-x_j}E_{ij}\;,\label{OCM131}
\end{eqnarray}
\end{subequations} 
where $x_i(n,m)$ is the position of the $i^{th}$ particle, $N$ is the number of particles in the system and $E_{ij}$ is the matrix with entries $(E_{ij})_{kl}=\delta_{ik}\delta_{jl}$. Here, $\wh x_i=x_i(m+1)$ is a forward shift and ${{\hypohat 0 x}_i}=x_i(m-1)$ is a backward shift.
\\
\\
\textbf{\emph{Discrete flow-$n$ direction}}: The compatibility between \eqref{OCM111} and \eqref{OCM121} gives us 
\begin{subequations} \label{OCM2} 
\begin{eqnarray}
\wt{\boldsymbol L_{\lambda}}\boldsymbol M&=&\boldsymbol M\boldsymbol L_{\lambda} \nn\\
\sum\limits_{i, j = 1}^N  \sum\limits_{k, \ell = 1}^N \frac{1}{(\wt{x}_i-\wt{x}_j+\lambda)} \frac{1}{(\wt{x}_k-x_\ell)} E_{ij} E_{k\ell}&=&\sum\limits_{i, j = 1}^N  \sum\limits_{k, \ell = 1}^N \frac{1}{(\wt{x}_i-x_j)} \frac{1}{(x_k-x_\ell+\lambda)} E_{ij} E_{k\ell}\;\nn\\\label{OCM21} 
\sum\limits_{i, \ell = 1}^N  \sum\limits_{k = 1}^N \frac{1}{(\wt{x}_i-\wt{x}_k+\lambda)(\wt{x}_k-x_\ell)} E_{i\ell}&=&\sum\limits_{i, \ell = 1}^N  \sum\limits_{k = 1}^N \frac{1}{(\wt{x}_i-x_k)(x_k-x_\ell+\lambda)} E_{k\ell}\;. \label{OCM22} 
\end{eqnarray}
Taking out a common factor, we obtain
\begin{equation}
\sum\limits_{k = 1}^N \left(\frac{1}{\wt{x}_i-\wt{x}_k+\lambda}-\frac{1}{\wt{x}_i-x_k} \right) =\sum\limits_{k = 1}^N \left(\frac{1}{x_k-x_\ell+\lambda}-\frac{1}{\wt{x}_k-x_\ell} \right) \;. \label{OCM23} 
\end{equation}
We see that both sides of \eqref{OCM23} are independent and, therefore, it holds if
\begin{equation}
\sum\limits_{k = 1}^N \left(\frac{1}{\wt{x}_i-\wt{x}_k+\lambda}-\frac{1}{\wt{x}_i-x_k} \right) \equiv \wt{p} \;, \label{OCM24} 
\end{equation}
where $p= p(n)$ is independent particle indices and a function of discrete time variable $n$. Taking a backward shift on \eqref{OCM24}, we have
\begin{equation}
\sum\limits_{k = 1}^N \left(\frac{1}{x_i-x_k+\lambda}-\frac{1}{x_i-{\hypotilde 0 x}_k} \right) = p \;. \label{OCM25} 
\end{equation}
Automatically, on the right-hand side of \eqref{OCM23}, we have
\begin{equation}
p= \sum\limits_{k = 1}^N \left(\frac{1}{x_\ell-\wt{x}_k}-\frac{1}{x_\ell-x_k-\lambda} \right)  \;. \label{OCM26} 
\end{equation}
Now, it is not difficult to see that, from \eqref{OCM25} and \eqref{OCM26}, we obtain
\begin{equation}
\sum\limits_{k = 1}^N \left(\frac{1}{x_i-\wt{x}_k}+\frac{1}{x_i-{\hypotilde 0 x}_k} \right) - \sum\limits_{k = 1}^N \left(\frac{1}{x_i-x_k+\lambda}-\frac{1}{x_i-x_k-\lambda} \right) = 0  \;, \label{OCM27} 
\end{equation}
which will be treated as a one-parameter discrete-time CM system and, under the limit: $\lambda \to 0$, one obtains
\begin{equation}
\sum\limits_{k = 1}^N \left(\frac{1}{x_i-\wt{x}_k}+\frac{1}{x_i-{\hypotilde 0 x}_k} \right) - \sum\limits_{k = 1 \atop k \neq i }^N \frac{2}{x_i-x_k}  = 0 \;, \label{OCM28} 
\end{equation}
\end{subequations} 
which is nothing but a standard discrete-time CM system in the $n$ direction.
\\
\\
\textbf{\emph{Discrete flow-$m$ direction}}: The compatibility between \eqref{OCM111} and \eqref{OCM131} gives us 
\begin{subequations} \label{OCM3} 

\begin{eqnarray}
\wh{\boldsymbol L_{\lambda}}\boldsymbol M&=&\boldsymbol M\boldsymbol L_{\lambda} \nn\\
\sum\limits_{i, j = 1}^N  \sum\limits_{k, \ell = 1}^N \frac{1}{(\wh{x}_i-\wh{x}_j+\lambda)} \frac{1}{(\wh{x}_k-x_\ell)} E_{ij} E_{k\ell}&=&\sum\limits_{i, j = 1}^N  \sum\limits_{k, \ell = 1}^N \frac{1}{(\wh{x}_i-x_j)} \frac{1}{(x_k-x_\ell+\lambda)} E_{ij} E_{k\ell}\;\nn\\\label{OCM31} 
\sum\limits_{i, \ell = 1}^N  \sum\limits_{k = 1}^N \frac{1}{(\wh{x}_i-\wh{x}_k+\lambda)(\wh{x}_k-x_\ell)} E_{i\ell}&=&\sum\limits_{i, \ell = 1}^N  \sum\limits_{k = 1}^N \frac{1}{(\wh{x}_i-x_k)(x_k-x_\ell+\lambda)} E_{k\ell}\;. \label{OCM32} 
\end{eqnarray}
Again, taking out a common factor, we obtain
\begin{equation}
\sum\limits_{k = 1}^N \left(\frac{1}{\wh{x}_i-\wh{x}_k+\lambda}-\frac{1}{\wh{x}_i-x_k} \right) =\sum\limits_{k = 1}^N \left(\frac{1}{x_k-x_\ell+\lambda}-\frac{1}{\wh{x}_k-x_\ell} \right) \;. \label{OCM33} 
\end{equation}
The situation is similar to the previous discrete flow. Both sides of \eqref{OCM33} are independent and it holds if 
\begin{equation}
\sum\limits_{k = 1}^N \left(\frac{1}{\wh{x}_i-\wh{x}_k+\lambda}-\frac{1}{\wh{x}_i-x_k} \right) \equiv \wh{q} \;, \label{OCM34} 
\end{equation}
where $q= q(m)$ is independent particle indices and a function of discrete time variable $m$. Computing a backward shift on \eqref{OCM34}, we obtain
\begin{equation}
\sum\limits_{k = 1}^N \left(\frac{1}{x_i-x_k+\lambda}-\frac{1}{x_i-{\hypohat 0 x}_k} \right) = q \;. \label{OCM35} 
\end{equation}
From the right-hand side of \eqref{OCM33}, we shall have
\begin{equation}
q= \sum\limits_{k = 1}^N \left(\frac{1}{x_\ell-\wh{x}_k}-\frac{1}{x_\ell-x_k-\lambda} \right)  \;. \label{OCM36} 
\end{equation}
Therefore, \eqref{OCM35} and \eqref{OCM36} give
\begin{equation}
\sum\limits_{k = 1}^N \left(\frac{1}{x_i-\wh{x}_k}+\frac{1}{x_i-{\hypohat 0 x}_k} \right) - \sum\limits_{k = 1}^N \left(\frac{1}{x_i-x_k+\lambda}-\frac{1}{x_i-x_k-\lambda} \right) = 0  \;, \label{OCM37} 
\end{equation}
which will be treated as a one-parameter discrete-time CM system in the $m$-direction and, under the limit: $\lambda \to 0$, we obtain
\begin{equation}
\sum\limits_{k = 1}^N \left(\frac{1}{x_i-\wh{x}_k}+\frac{1}{x_i-{\hypohat 0 x}_k} \right) - \sum\limits_{k = 1 \atop k \neq i }^N \frac{2}{x_i-x_k}  = 0  \;, \label{OCM38} 
\end{equation}
\end{subequations} 
which is a discrete-time CM system in the $m$ direction.
\\
\\
\textbf{\emph{Commutativity between discrete flows}}: Two discrete-time dynamics will be consistent if the compatibility between \eqref{OCM121} and \eqref{OCM131} 
\begin{subequations}
\begin{eqnarray}\label{OCM4}
\wh{\boldsymbol M}\boldsymbol N&=&\wt{\boldsymbol N}\boldsymbol M \;
\end{eqnarray}
holds. This gives us a set of equations
\begin{equation}
p-q= \sum\limits_{k = 1}^N \left(\frac{1}{x_i-\wt{x}_k}-\frac{1}{x_i-\wh{x}_k} \right)  \;, \label{OCM41}
\end{equation}
and 
\begin{equation}
p-q= \sum\limits_{k = 1}^N \left(\frac{1}{x_i-{\hypotilde 0 x}_k}-\frac{1}{x_i-{\hypohat 0 x}_k} \right)  \;, \label{OCM42}
\end{equation}
which will be called corner equations.
Imposing \eqref{OCM41} = \eqref{OCM42}, we obtain
\begin{equation}
\sum\limits_{k = 1}^N \left(\frac{1}{x_i-\wt{x}_k}+\frac{1}{x_i-{\hypotilde 0 x}_k} \right) = \sum\limits_{k = 1}^N \left(\frac{1}{x_i-\wh{x}_k}-\frac{1}{x_i-{\hypohat 0 x}_k} \right)  \;, \label{OCM43}
\end{equation}
\end{subequations} 
which is a constraint equation given the connection between one discrete flow and another discrete flow.
%%%%%%%%%%%%%%%%%%%%%%%%%%%%%%%%%%%%%%%%%%%%%%%%%%%%%%%%%%%
%%%%%%%%%%%%%%%%%%%%%%%%%%%%%%%%%%%%%%%%%%%%%%%%%%%%%%%%%%%%%
\section{Integrability: the closure relation }
\setcounter{equation}{0} 
In this section, we will show that the one-parameter discrete-time CM systems in the previous section are integrable in the sense that their discrete-time Lagrangians satisfy the closure relation as a consequence of the least action principle with respect to the independent variables \cite{LNQ, LooN, YooNij, YooNij1, JaiYoo, JaiYoo1, PieYoo, BooSur}.
\\
\\
It is not difficult to see that \eqref{OCM27} and \eqref{OCM37} can be obtained from the discrete Euler-Lagrange equations \cite{YooNij}
\begin{eqnarray}
\wt{\frac{\partial\mathcal{L}_n(x, \wt{x})}{\partial x_i}}+ \frac{\partial\mathcal{L}_n(x, \wt{x})}{\partial{\wt{x}_i}}=0\;,\label{OCM53}
\end{eqnarray}
\begin{eqnarray}
\wh{\frac{\partial\mathcal{L}_m(x, \wh{x})}{\partial x_i}}+ \frac{\partial\mathcal{L}_m(x, \wh{x})}{\partial{\wh{x}_i}}=0\;,\label{OCM54}
\end{eqnarray}
where
%\begin{subequations} \label{OCM5}
\begin{eqnarray}
\mathcal{L}_n{(x, \wt{x})}=-\sum\limits_{i,j=1}^N \ln \left | x_i- \wt{x}_j \right | +\sum\limits_{i,j=1}^N  \ln \left | x_i - x_j + \lambda \right | +p(\Xi -\wt{\Xi })\;,\label{OCM51}
\end{eqnarray}
and
\begin{eqnarray}
\mathcal{L}_m{(x, \wh{x})}=-\sum\limits_{i,j=1}^N \ln \left | x_i- \wh{x}_j \right | +\sum\limits_{i,j=1}^N \ln \left | x_i - x_j + \lambda \right |+q(\Xi -\wh{\Xi })\;.\label{OCM52}
\end{eqnarray}
%\end{subequations}
Here $\Xi =\sum\limits_{i=1}^Nx_i$ is a centre of mass variable.
\\
\\
To show that the Lagrangian closure relation for the one-parameter discrete-time CM model holds, we shall employ a connection between the temporal Lax matrix and Lagrangian as we did have in the case of the standard discrete-time CM model \cite{YooNij}. An interesting point is that, for this present system, it turns out that one could obtain the discrete-time Lagrangian from the relation $\mathcal{L}(x, \wt{x}) = \ln \left | \det \boldsymbol M_{RS} \right |$\footnote{See the appendix A  for the explicit computation.}, where $\boldsymbol M_{RS}$ is a temporal matrix for the RS model given by
\begin{eqnarray}
\boldsymbol M_{RS}=\sum\limits_{i,j=1}^N\frac{\wt{h}_ih_j}{\wt{x}_i-x_j+\lambda}E_{ij}\;,\label{OCM6}
\end{eqnarray}
where $h_i=h_i(n,m)$ are auxiliary variables which can be determined \cite{Nij-R-K}. Suppose there is another temporal matrix given by
%begin{subequations} \label{OCM7}
\begin{eqnarray}
\boldsymbol N_{RS} = \sum\limits_{i,j=1}^N \frac{\wh{h}_ih_j}{\wh{x}_i-x_j+\lambda}E_{ij}\;, \label{OCM72}
\end{eqnarray}
and both $\boldsymbol{M}_{RS}$ and $\boldsymbol{N}_{RS}$ satisfy 
\begin{eqnarray}
\wh{\boldsymbol M}_{RS} \boldsymbol N_{RS}  =\wt{\boldsymbol N}_{RS}\boldsymbol M_{RS}.\; \label{OCM77}
\end{eqnarray}
Taking $\det$ and $\ln$, one obtains 
\begin{eqnarray}
\ln \left | \det \wh{\boldsymbol M}_{RS} \right | + \ln \left | \det \boldsymbol N_{RS} \right | =\ln \left | \det \wt{\boldsymbol N}_{RS} \right |+\ln \left | \det \boldsymbol M_{RS} \right |\;,
\end{eqnarray}
resulting in the closure relation \eqref{OCM78}.
%\begin{equation}\label{OCM78}
% \widehat{\mathcal{L}(\boldsymbol{x},\widetilde{\boldsymbol{x}})} - \mathcal{L}(\boldsymbol{x},\widetilde{\boldsymbol{x}}) -
 %\widetilde{\mathcal{L}(\boldsymbol{x},\widehat{\boldsymbol{x}})} + \mathcal{L}(\boldsymbol{x},\widehat{\boldsymbol{x}}) = 0\;,
%\end{equation}
%which is nothing but the closure relation.
%\end{subequations} 
%%%%%%%%%%%%%%%%%%%%%%%%%%%%%%%%%%%%%%%%%%%%%%%%%%%%%%%%%%
%%%%%%%%%%%%%%%%%%%%%%%%%%%%%%%%%%%%%%%%%%%%%%%%%%%%%%%%%%%%%
\section{Integrability: the classical r-matrix}
In this section, we shall construct the classical r-matrix for the one-parameter discrete-time CM system. We first rewrite the spatial Lax matrix as
\begin{eqnarray}
\boldsymbol L_\lambda&=&\sum\limits_{i=1}^N \frac{1}{\lambda} E_{ii} -\sum\limits_{i,j=1  \atop j \neq i }^N\frac{1}{x_i-x_j+\lambda}E_{ij}\;.\label{OCM143}
\end{eqnarray}
Next, we shall call the spatial Lax matrix of the standard CM system \cite{AvaTal} given by
\begin{eqnarray}
\boldsymbol L=\sum\limits_{i=1}^N P_i E_{ii} -\sum\limits_{i,j=1  \atop k \neq i }^N\frac{1}{x_i-x_j}E_{ij}\;,\label{OCM141}
\end{eqnarray}
where $P_i$ is the momentum variable for $i^{th}$ particle. With this structure, one finds that the classical r-matrix can be computed through the relation
\begin{eqnarray}
\{\boldsymbol L \overset{\otimes}{,} \boldsymbol L\} &=& [r_{12}, \boldsymbol L \otimes \mathds{1}]-[r_{12},\mathds{1} \otimes \boldsymbol L]\;,\label{OCM142}
\end{eqnarray}
where $r_{12}$ is the classical r-matrix for the CM system. Comparing \eqref{OCM143} with \eqref{OCM141}, one immediately finds the classical r-matrix $r_{12}^\lambda$ for the one-parameter discrete-time CM system upon replacing $P_i \to \frac{1}{\lambda}$ and $\frac{1}{x_i-x_j} \to \frac{1}{x_i-x_j+\lambda}$
\begin{eqnarray}\label{OCM}
\{\boldsymbol L_\lambda \overset{\otimes}{,} \boldsymbol L_\lambda \}&=&\left[r_{12}^\lambda, \boldsymbol L_\lambda \otimes  \mathds{1} \right]- \left[r_{12}^\lambda,  \mathds{1} \otimes \boldsymbol L_\lambda\right]\;.
\end{eqnarray}
 We shall note here that under the limit $\lambda\to 0$, the classical r-matrix $r_{12}^\lambda$ will not yield the standard classical r-matrix. This problem arises from the fact that the spatial Lax matrix \eqref{OCM143} is a fake one since it does not provide the integrals of motion through the relation $I_n=\frac{1}{n!}Tr \boldsymbol{L_\lambda}^n$. 
%%%%%%%%%%%%%%%%%%%%%%%%%%%%%%%%%%%%%%%%%%%%%%%%%%%%%%%%%%%%%%%%%%%%%%%%%%%%%%%%%%%
%%%%%%%%%%%%%%%%%%%%%%%%%%%%%%%%%%%%%%%%%%%%%%%%%%%%%%%%%%%%%%%%%%%%%%%%%%%%%%%%%%%
\section{Integrability: the exact solution} \label{ES}
\begin{subequations}\label{OCM8} 
In this section, we will construct the exact solution $\{x_i(n)\}$ with initial values $\{x_i(0)\}$ and $\{x_i(1)=\wtilde x_i(0)\}$. We shall first rewrite the Lax matrices in the forms
%\begin{eqnarray}
%\boldsymbol L&=&\sum_{i,j=1}^N\frac{1}{x_i-x_j+\lambda}E_{ij}\;,\label{OCM81}
%\end{eqnarray}
\begin{eqnarray}
\boldsymbol X \boldsymbol L - \boldsymbol L \boldsymbol X + \lambda \boldsymbol L =  \boldsymbol E, \;\label{OCM82}
\end{eqnarray}
%\begin{eqnarray}
%\boldsymbol M&=&\sum_{i,j=1}^N\frac{1}{\wt{x}_i-x_j}E_{ij}\;,\label{OCM83}
%\end{eqnarray}
\begin{eqnarray}
\wt{\boldsymbol X}\boldsymbol{M}-\boldsymbol{M}\boldsymbol X=\boldsymbol E \;,\label{OCM84}
\end{eqnarray}
where $\boldsymbol X  =  \sum\limits_{i=1}^N x_i E_{ii}$ and $\boldsymbol E = \sum\limits_{i=1}^N  E_{ij}$.
Moreover, we have 
\begin{eqnarray}
(\wt{\boldsymbol{L}}-\boldsymbol{M}) \boldsymbol{E} = 0 \;,\label{OCM85}
\end{eqnarray}
and 
\begin{eqnarray}
\boldsymbol{E} (\boldsymbol{L}-\boldsymbol{M}) = 0 \;,\label{OCM86}
\end{eqnarray}
with also give equations of motion. Let's write $\boldsymbol{M} = \wt{\boldsymbol U}\boldsymbol U^{-1}$ and $\boldsymbol{L} =\boldsymbol U\bLam \boldsymbol U^{-1}$, where $\boldsymbol U$ is an invertible matrix.
\eqref{OCM84} leads to 
\begin{eqnarray}
\wt{\boldsymbol X} \wt{\boldsymbol U} \boldsymbol U^{-1} -\wt{\boldsymbol U} \boldsymbol U^{-1}\boldsymbol X &=& \boldsymbol{E} \;\nn\\
\boldsymbol U^{-1} \wt{\boldsymbol X}\wt{\boldsymbol U} \boldsymbol U^{-1} -\wt{\boldsymbol U}^{-1}\wt{\boldsymbol U} \boldsymbol U^{-1}\boldsymbol X &=&\wt{\boldsymbol U}^{-1} \boldsymbol{E} \;\nn\\
\wt{\boldsymbol U}^{-1} \wt{\boldsymbol X}\wt{\boldsymbol U} \boldsymbol U^{-1}\boldsymbol U - \boldsymbol U^{-1}\boldsymbol X \boldsymbol U&=&\wt{\boldsymbol U}^{-1} \boldsymbol{E}\boldsymbol U \;\nn\\
\wt{\boldsymbol U}^{-1} \wt{\boldsymbol X}\wt{\boldsymbol U}  - \boldsymbol U^{-1}\boldsymbol X\boldsymbol U &=&\wt{\boldsymbol U}^{-1} \boldsymbol{E}\boldsymbol U \;\nn\\
\wt{\boldsymbol Y} - \boldsymbol Y &=&\wt{\boldsymbol U}^{-1} \boldsymbol{E}\boldsymbol U,  \;\label{OCM87}
\end{eqnarray}
where $ \boldsymbol Y =\boldsymbol U^{-1} \boldsymbol{X}\boldsymbol U$. We also find that \eqref{OCM82} gives 
\begin{eqnarray}
\boldsymbol X \boldsymbol U\bLam \boldsymbol U^{-1}-\boldsymbol U\bLam \boldsymbol U^{-1}\boldsymbol X + \lambda \boldsymbol U\bLam \boldsymbol U^{-1}&=& \boldsymbol{E} \;\nn\\
\boldsymbol X \boldsymbol U\bLam \boldsymbol U^{-1} \boldsymbol U - \boldsymbol U\bLam \boldsymbol U^{-1}\boldsymbol X \boldsymbol U + \lambda \boldsymbol U \bLam\boldsymbol U^{-1} \boldsymbol U &=& \boldsymbol{E} \boldsymbol U \;\nn\\
\boldsymbol U^{-1}\boldsymbol X \boldsymbol U\bLam - \boldsymbol U^{-1}\boldsymbol U\bLam \boldsymbol U^{-1}\boldsymbol X \boldsymbol U + \boldsymbol U^{-1}\lambda \boldsymbol U \bLam &=&\boldsymbol U^{-1} \boldsymbol{E} \boldsymbol U \;\nn\\
\boldsymbol U^{-1}\boldsymbol X \boldsymbol U\bLam - \bLam \boldsymbol U^{-1}\boldsymbol X \boldsymbol U + \lambda \boldsymbol U^{-1} \boldsymbol U\bLam &=&\boldsymbol U^{-1} \boldsymbol{E} \boldsymbol U \;\nn\\
\boldsymbol U^{-1}\boldsymbol X \boldsymbol U\bLam - \bLam \boldsymbol U^{-1}\boldsymbol X \boldsymbol U + \lambda \bLam&=&\boldsymbol U^{-1} \boldsymbol{E} \boldsymbol U \;\nn\\
\boldsymbol Y \bLam - \bLam \boldsymbol Y+ \lambda \bLam  &=& \boldsymbol U^{-1} \boldsymbol{E} \boldsymbol U \;, \label{OCM88}
\end{eqnarray}
and \eqref{OCM85} gives
\begin{eqnarray}
\left( \wt{\boldsymbol U}\bLam \wt{\boldsymbol U}^{-1}-\wt{\boldsymbol U}\boldsymbol U^{-1} \right) \boldsymbol E &=& 0 \;\nn\\
\wt{\boldsymbol U}\bLam \wt{\boldsymbol U}^{-1}\boldsymbol E-\wt{\boldsymbol U}\boldsymbol U^{-1}\boldsymbol E &=& 0 \;\nn\\
\wt{\boldsymbol U}^{-1} \wt{\boldsymbol U}\bLam \wt{\boldsymbol U}^{-1}\boldsymbol E-\wt{\boldsymbol U}^{-1}\wt{\boldsymbol U}\boldsymbol U^{-1}\boldsymbol E &=& 0 \;\nn\\
\bLam \wt{\boldsymbol U}^{-1}\boldsymbol E - \boldsymbol U^{-1}\boldsymbol E &=& 0 \;\nn\\\boldsymbol U^{-1}\boldsymbol E\boldsymbol U &=& \bLam \wt{\boldsymbol U}^{-1}\boldsymbol E\boldsymbol U \;.\label{OCM89}
\end{eqnarray}
Substituting \eqref{OCM89} into \eqref{OCM88}, we obtain
\begin{eqnarray}
\boldsymbol Y \bLam - \bLam \boldsymbol Y+ \lambda \bLam  &=& \bLam \wt{\boldsymbol U}^{-1}\boldsymbol E\boldsymbol U \;.\label{OCM810}
\end{eqnarray}
To eliminate the invertible matrix $\boldsymbol{U}$ and $\boldsymbol{E}$ on the right hand side of \eqref{OCM810}, we use \eqref{OCM86} which can be expressed in the form
\begin{eqnarray}
 \boldsymbol E \left(\boldsymbol U \bLam \boldsymbol {U}^{-1} - \wt{\boldsymbol U}\boldsymbol U^{-1} \right) &=& 0 \;\nn\\
 \boldsymbol E \boldsymbol U \bLam \boldsymbol {U}^{-1} - \boldsymbol E\wt{\boldsymbol U}\boldsymbol U^{-1} &=& 0 \;\nn\\
\boldsymbol E \boldsymbol U \bLam \boldsymbol{U}^{-1}\boldsymbol U - \boldsymbol E\wt{\boldsymbol U}\boldsymbol U^{-1}\boldsymbol U &=& 0 \;\nn\\
\boldsymbol E \boldsymbol U \bLam  - \boldsymbol E\wt{\boldsymbol U} &=& 0 \;\nn\\
  \boldsymbol U^{-1} \boldsymbol E\wt{\boldsymbol U} &=&\boldsymbol U^{-1} \boldsymbol E \boldsymbol U \bLam \;. \label{OCM811}
 \end{eqnarray}
Since $\boldsymbol U^{-1} \boldsymbol E\wt{\boldsymbol U} = \wt{\boldsymbol U}^{-1} \boldsymbol E \boldsymbol U$, we then obtain
\begin{eqnarray}
\wt{\boldsymbol U}^{-1} \boldsymbol E \boldsymbol U&=& \boldsymbol U^{-1} \boldsymbol E \boldsymbol U \bLam \;.\label{OCM812}
\end{eqnarray}
\end{subequations} 
Substituting \eqref{OCM811} into \eqref{OCM87}, one finds
\begin{eqnarray}
\wt{\boldsymbol Y} - \boldsymbol Y &=&\boldsymbol U^{-1} \boldsymbol{E}\boldsymbol U\bLam\;.  \;\label{OCM9}
\end{eqnarray}
Rearranging \eqref{OCM810}, we obtain
\begin{eqnarray}
\bLam^{-1}\boldsymbol Y\bLam -\bLam^{-1}\bLam \boldsymbol Y+ \bLam^{-1}\lambda \bLam &=&\bLam^{-1}\bLam \wt{\boldsymbol U}^{-1} \boldsymbol{E}\boldsymbol U,\;\nn\\
\bLam^{-1}\boldsymbol Y\bLam - \boldsymbol Y +\lambda &=& \wt{\boldsymbol U}^{-1} \boldsymbol{E}\boldsymbol U \;.\label{OCM10}
\end{eqnarray}
Substituting \eqref{OCM87} into \eqref{OCM10}, we get
\begin{eqnarray}
\bLam^{-1}\boldsymbol Y\bLam - \boldsymbol Y +\lambda &=& \wt{\boldsymbol Y} \boldsymbol Y  ,\;\nn\\
\wt{\boldsymbol Y}  &=&\bLam^{-1} \boldsymbol Y\bLam +  \lambda \;.\label{OCM11}
 \end{eqnarray}
Hence if we proceed $n$ steps, we find that
\begin{eqnarray}
\wt{\wt{\boldsymbol Y}}  &=&\bLam^{-1}\wt{\boldsymbol Y}\bLam +  \lambda ,\;\nn\\
 &=&\bLam^{-1} \left[\bLam^{-1}\boldsymbol Y\bLam +  \lambda\right]\bLam  + \lambda, \;\nn\\
 &=&\left(\bLam^{-1}\right)^{2} \boldsymbol Y \bLam^2 +  2\lambda , \;\nn\\
 &&.\;\nn\\
 &&.\;\nn\\
 &&.\;\nn\\
\boldsymbol Y(n)&=& \left(\bLam\right)^{-n} \boldsymbol Y \bLam^n +  n\lambda \;\label{OCM11}
 \end{eqnarray}
and, of course, for the $m$ steps
\begin{eqnarray}
\boldsymbol Y(m)&=& \left(\bLam\right)^{-m} \boldsymbol Y \bLam^m +  m\lambda .\;\label{OCM12}
 \end{eqnarray}
Then, at any $(n,m)$ steps, we have 
\begin{eqnarray}
\boldsymbol Y(n,m)&=& \left(p+\bLam \right)^{-n} \left(q+\bLam \right)^{-m}\boldsymbol Y(0,0) \left(q+\bLam \right)^{m}\left(p+\bLam \right)^{n}+  (n+m)\lambda .\;\label{OCM13}
 \end{eqnarray}
It is not difficult to find that, under the limit $\lambda \mapsto 0$, one obtains
\begin{eqnarray}
\boldsymbol Y(n,m)&=& \left(p+\bLam \right)^{-n} \left(q+\bLam \right)^{-m}\boldsymbol Y(0,0) \left(q+\bLam \right)^{m}\left(p+\bLam \right)^{n} ,\;\label{OCM1311}
 \end{eqnarray}
 which is nothing but a standard solution of the discrete-time CM system \cite{Nij-Pan}.
%%%%%%%%%%%%%%%%%%%%%%%%%%%%%%%%%%%%%%%%%%%%%%%%%%%%%%%%%%%%%%%%%%%%%%%%%%%%%%%%%%%
\section{The continuum limit}\label{CL}
\setcounter{equation}{0} 
In this section, we consider the continuum limit of the one-parameter discrete-time CM system which had been investigated in the previous sections. Since there are two discrete-time variables $(n,m)$, we may perform a naive continuum limit \cite{Nij-Pan} on each of these variables resulting in the one-parameter continuous-time CM system. To proceed such continuuum limit, we define $x_i = Z_i + n\Delta$, where $\Delta$ is a small parameter. Consequently, we also have   $\wt{x}_i = \wt{Z}_i + (n+1)\Delta$ and ${\hypotilde 0 x}_i = {\hypotilde 0 Z}_i + (n-1)\Delta$. Then
\eqref{OCM27} becomes
\begin{eqnarray}
\sum\limits_{k=1}^N \left(\frac{1}{Z_i-\wt{Z}_k-\Delta} + \frac{1}{Z_i-{\hypotilde 0 Z}_k+\Delta}\right) -
\sum\limits_{i,k=1 \atop k \neq i }^N \left(\frac{1}{Z_i - Z_k+\lambda} + \frac{1}{Z_i - Z_k-\lambda}\right) = 0 \;.\label{OCM151}
\end{eqnarray}
or
\begin{eqnarray}
&& \left(\frac{1}{Z_i-\wt{Z}_i-\Delta} + \frac{1}{Z_i-{\hypotilde 0 Z}_i+\Delta}\right) - \nn\\
&&\sum\limits_{i,k=1 \atop k \neq i }^N \left(\frac{1}{Z_i - \wt{Z}_k-\Delta} + \frac{1}{Z_i - {\hypotilde 0 Z}_k+\Delta} -  \frac{1}{Z_i - Z_k+\lambda} -  \frac{1}{Z_i - Z_k-\lambda}\right) =0\;. \; \label{OCM151}
\end{eqnarray}
Taking the expansion, we get
\begin{eqnarray}
\wt{ Z}_i &=& Z_i+ \varepsilon\frac{d Z_i}{dt}+\frac{\varepsilon^2}{2}\frac{d^2 Z_i}{dt^2} + ...\;,\\
{\hypotilde 0 Z}_i &=& Z_i- \varepsilon\frac{d Z_i}{dt}+\frac{\varepsilon^2}{2}\frac{d^2 Z_i}{dt^2} + ...\;, \label{OCM152}
\end{eqnarray}
where $\varepsilon$ is the time-step parameter. Then, the first two terms in \eqref {OCM151} can be expressed in the form
\begin{eqnarray}
\left(\frac{1}{Z_i-\wt{Z}_i-\Delta} + \frac{1}{Z_i-{\hypotilde 0 Z}_i+\Delta}\right)&=&\frac{\varepsilon^2}{\Delta^2}\frac{d^2 Z_i}{dt^2}+... \;. \label{OCM153}
\end{eqnarray}
We also find that
\begin{eqnarray}
&&\sum\limits_{k=1 \atop k \neq i }^N \left(\frac{1}{Z_i - \wt{Z}_k-\Delta} + \frac{1}{Z_i - {\hypotilde 0 Z}_k+\Delta}\right) \nn\\ 
%&&= \sum\limits_{k=1 \atop k \neq i }^N\left(\frac{2}{Z_i-Z_k} %+ \frac{1}{(Z_i-Z_k)^3}\left( \varepsilon^2 \left(\frac{dZ_k}%{dt}\right)^2 + 2 \varepsilon \Delta \frac{dZ_k}{dt} + \Delta^2 %\right) \nn\\
&&= \sum\limits_{k=1 \atop k \neq i }^N\left(\frac{2}{Z_i-Z_k} + \frac{1}{(Z_i-Z_k)^3}\left( \varepsilon^2 \frac{dZ_k}{dt} + 2 \varepsilon \Delta \frac{dZ_k}{dt} + \Delta^2 \right)+....\right) \;.\label{OCM154}
\end{eqnarray}
If $\varepsilon \approx \Delta^2$, one finds that
\begin{eqnarray}
\sum\limits_{k=1 \atop k \neq i }^N \left(\frac{1}{Z_i - \wt{Z}_k-\Delta} + \frac{1}{Z_i - {\hypotilde 0 Z}_k+\Delta}\right)&\approx& \sum\limits_{k=1 \atop k \neq i }^N\left(\frac{2}{Z_i-Z_k} + \frac{2 \Delta^2}{(Z_i-Z_k)^3}\right) \;. \label{OCM155}
\end{eqnarray}
Finally, the continuous version of the one-parameter CM system is given by
\begin{eqnarray}
\frac{d^2 Z_i}{dt^2} + \sum\limits_{i,k=1 \atop k \neq i }^N \left( g^{\prime}\left[ \frac{2}{Z_i-Z_k} - \frac{1}{Z_i-Z_k + \lambda}-\frac{1}{Z_i-Z_k - \lambda} \right] + \frac{2g}{(Z_i-Z_k)^3} \right) = 0   \;, \label{OCM156}
\end{eqnarray}
where $g \equiv \frac{\Delta^4}{\varepsilon^2}$ and $g^{\prime} \equiv \frac{\Delta^2}{\varepsilon^2}$.
Therefore, under the limit: $\lambda \to 0$, we have
\begin{eqnarray}
\frac{d^2 Z_i}{dt^2} + 2g \sum\limits_{i,k=1 \atop k \neq i }^N  \frac{1}{(Z_i-Z_k)^3} &=& 0   \;, \label{OCM157}
\end{eqnarray}
which is actually a standard continuous CM system.
With \eqref{OCM156}, 
the Lagrangian is given by
\begin{eqnarray}\label{OCM158}
\mathscr L_\lambda=\sum\limits_{i=1}^N \frac{\partial Z_i}{\partial t}-\frac{1}{2}\sum\limits_{i,k=1 \atop k \neq i }^N \frac{g}{(Z_i-Z_k)^2}-g^{\prime} \sum\limits_{i,k=1 \atop k \neq i }^N \left(\ln \left|Z_i-Z_k+ \lambda\right|+\ln\left|(Z_i-Z_k)\right|\right)
\end{eqnarray} 
with the Euler-Lagrange equation 
\begin{eqnarray}\label{OCM159}
\frac{\partial \mathscr L_{\lambda}}{\partial Z_i}-\frac{\partial}{\partial t}\left( \frac{\partial \mathscr 
L_{\lambda}}{\partial(\frac{\partial Z_i}{\partial t})}\right)=0\;. 
\end{eqnarray}
Of course, under the limit $\lambda \to 0$, 
\begin{equation}
    \lim_{\lambda\mapsto 0}\mathscr L_\lambda=\mathscr L=\sum\limits_{i=1}^N \frac{\partial Z_i}{\partial t}+\sum\limits_{i,k=1 \atop k \neq i }^N \frac{g}{(Z_i-Z_k)^2}\;,
\end{equation}
the standard Lagrangian for the CM system is recovered\footnote{We note that the CM system in this equation comes with the opposite sign with the standard one.}.
In addition, the Hamiltonian of the one-parameter continuous-time of CM system can be written in the form 
\begin{eqnarray}\label{OCM160}
\mathscr H_\lambda = \sum\limits_{i=1}^N P_i^2+\frac{1}{2}\sum\limits_{i,k=1 \atop k \neq i }^N \frac{g}{(Z_i-Z_k)^2}+g^{\prime} \sum\limits_{i,k=1 \atop k \neq i }^N \left(\ln \left|Z_i-Z_k+ \lambda\right|+\ln\left|Z_i-Z_k\right|\right),\;
\end{eqnarray}
where $P_i = \frac{\partial Z_i}{\partial t}$ is the momentum variable for the $i^{th}$ particle.
%%%%%%%%%%%%%%%%%%%%%%%%%%%%%%%%%%%%%%%%%%%%%%%%%%%%%%%%%%%%%%%%%%%%%%%%%%%%%%%%
%%%%%%%%%%%%%%%%%%%%%%%%%%%%%%%%%%%%%%%%
%%%%%%%%%%%%%%%%%%%%%%%%%%%%%%%%%%%%%%%%
\section{Summary}
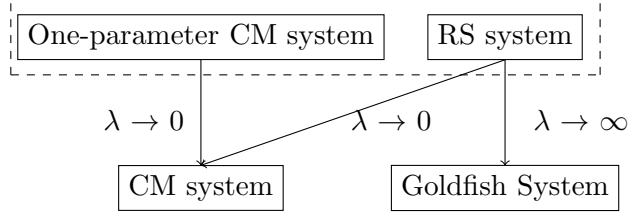
\begin{figure}[h]
\begin{center}
\begin{tikzpicture}
    \node (OP)    at (-1,0) [draw] {One-parameter CM system};
  
  \node (RS) at (3,0)  [draw] {RS system};
  
            \draw[->] (-1,-0.3) -- (-1,-1.7);     
            \draw[->] (3,-0.3) -- (3,-1.7);
          \draw[->] (3,-0.3) -- (-1,-1.7);
\draw (-1.75,-1.1) node {$\lambda$ $\to$ $0$}; 
\draw (1.5,-1.1) node {$\lambda$ $\to$ $0$}; 
\draw (4.0,-1.1) node {$\lambda$ $\to$ $\infty$}; 
   \node (CM)    at (-1,-2) [draw] {CM system};
  \node (t4) at (3,-2)  [draw] {Goldfish System};
          \draw [dashed] (-3.5,-0.5) -- (4.25,-0.5);
          \draw [dashed] (-3.5,-0.5) -- (-3.5,0.5); 
          \draw [dashed] (-3.5,0.5) -- (4.25,0.5); 
          \draw [dashed] (4.25,-0.5) -- (4.25,0.5); 
 \end{tikzpicture}
\end{center}
\caption{ The connection among one-parameter CM, RS, CM and Goldfish systems.}\label{F2}
\end{figure}

In the present work, we propose a new type of integrable one-dimensional many-body system called a one-parameter or a deformed discrete-time CM system. Under the limit: $\lambda \to 0$, a standard CM system is recovered in both discrete and continuous cases. In figure \ref{F2}, we provide a diagram of the connection for all CM-type systems. We would rank our model on the same level as the RS system since both systems contain a parameter.
%
%
%report the new system, called the one-parameter discrete CM system. Adding one parameter method, the two discrete-time one-parameter discrete CM system is obtained from the Lax pairs. The key relation called the closure relation is established through the connection between the discrete-time Lagrangian and the temporal Lax matrix.  The full limit is performed to obtain the one-parameter continuous-time CM system. Noticeable discovery, the reduction of the one-parameter in both the discrete and continuous case leads to the rational CM system. In addition, we discover that the discrete Lagrangian of this system can be obtained determinant of the M Matrix of the RS system. The one-parameter CM and RS systems are the same hierarchy as the same one parameter($\lambda$). The both reduce $\lambda$ to 0 we obtain the CM system. moreover we obtain the Goldfish system when $\lambda \to \infty$ in RS system as shown in \ref{F2}.
\\
\\
We also would like to note that the continuous system obtained in section \ref{CL} is just the first one in CM hierarchy \cite{YooNij}. A question can be addressed here is that ``how is the other one deformed in the hierarchy?" Moreover, one also can try to study the integrability condition as well as the quantum property of the system. Further investigation is needed and we shall answer these points elsewhere. 
\appendix
%%%%%%%%%%%%%%%%%%%%%%%%%%%%%%%%%%%%%%%%%%%%%%%%%%%%%%%%%%%%%%%%%%%%%%%%
\section{The connection between Lagrangian and $\boldsymbol M_{RS}$ of the RS model}
\numberwithin{equation}{section}
In this appendix, we will derive the connection between one-parameter discrete-time Lagrangian and $\boldsymbol M_{RS}$
\begin{eqnarray}
\boldsymbol M_{RS}=\sum\limits_{i,j=1}^N\frac{\wt{h}_ih_j}{\wt{x}_i-x_j+\lambda}E_{ij}\;.\label{OCM6}
\end{eqnarray}
For simplicity, we shall first start with the case of $2\times 2$ matrix given by
\[ \boldsymbol M_{RS} 
=
\begin{bmatrix}
    \frac{\wt{h}_1h_1}{\wt{x}_1-x_1+\lambda} &  \frac{\wt{h}_1h_2}{\wt{x}_1-x_2+\lambda}  \\
    \frac{\wt{h}_2h_1}{\wt{x}_2-x_1+\lambda}  &  \frac{\wt{h}_2h_2}{\wt{x}_2-x_2+\lambda}   
\end{bmatrix}
\;.\] 
Then, we compute the determinant
\begin{eqnarray}
\det \boldsymbol M_{RS} &=&\frac{\wt{h}_1h_1\wt{h}_2h_2}{(\wt{x}_1-x_1+\lambda)(\wt{x}_2-x_2+\lambda)}- \frac{\wt{h}_2h_1\wt{h}_1h_2}{(\wt{x}_2-x_1+\lambda)(\wt{x}_1-x_2+\lambda)},\;\nn\\
&=& h_1\wt{h}_1h_2\wt{h}_2 \left[ \frac{1}{(\wt{x}_1-x_1+\lambda)(\wt{x}_2-x_2+\lambda)} - \frac{1}{(\wt{x}_2-x_1+\lambda)(\wt{x}_1-x_2+\lambda)} \right],\; \nn\\
&=&h_1\wt{h}_1h_2\wt{h}_2 \left[ \frac{(\wt{x}_2-x_1+\lambda)(\wt{x}_1-x_2+\lambda) - (\wt{x}_1-x_1+\lambda)(\wt{x}_2-x_2+\lambda)}{\prod\limits_{i,j=1}^2 (\wt{x}_i-x_j+\lambda)} \right].\; \nn\\ \label {AOCM}
\end{eqnarray}
\eqref{AOCM} can be further simplified as follows
\begin{eqnarray}
\det \boldsymbol M_{RS} &=& h_1\wt{h}_1h_2\wt{h}_2 \left[ \frac{(\wt{x}_2-\wt{x}_1)(x_1-x_2)}{\prod\limits_{i,j=1}^2 (\wt{x}_i-x_j+\lambda)}\right]\;. \label {AOCM1}
\end{eqnarray}
Recalling the relations \cite{YooNij1}
\begin{eqnarray}
h^{2}_i&=&-\frac{\prod\limits_{j=1}^N (x_i-x_j+\lambda)(x_i-\wt{x}_j-\lambda)}{\prod\limits_{\mathop {i,j = 1}\limits_{j \ne i} }^N(x_i-x_j)\prod\limits_{j=1}^N (x_i-\wt{x}_j)},\; \\
{\wt{h}_i}^2&=&-\frac{\prod\limits_{j=1}^N (\wt{x}_i-x_j+\lambda)(\wt{x}_i-\wt{x}_j-\lambda)}{\prod\limits_{\mathop {i,j = 1}\limits_{j \ne i} }^N(\wt{x}_i-\wt{x}_j)\prod\limits_{j=1}^N (\wt{x}_i-x_j)} \;,
\end{eqnarray}
then, for $i,j = 1,2$, we have
\begin{eqnarray}
h^{2}_1&=&-\frac{(x_1-x_1+\lambda)(x_1-x_2+\lambda)(x_1-\wt{x}_1-\lambda)(x_1-\wt{x}_2-\lambda)}{(x_1-x_2)(x_1-\wt{x}_1)(x_1-\wt{x}_2)},\; \\
{\wt{h}_1}^{2}&=&\frac{(\wt{x}_1-x_1+\lambda)(\wh{x}_1-x_2+\lambda)(\wt{x}_1-\wt{x}_1-\lambda)(\wt{x}_1-\wt{x}_2-\lambda)}{(\wt{x}_1-\wt{x}_2)(\wt{x}_1-x_1)(\wt{x}_1-x_2)},\; \\
h^{2}_1&=&-\frac{(x_2-x_1+\lambda)(x_2-x_2+\lambda)(x_2-\wt{x}_1-\lambda)(x_2-\wt{x}_2-\lambda)}{(x_2-x_1)(x_2-\wt{x}_1)(x_2-\wt{x}_2)},\; \\
{\wt{h}_1}^{2}&=&\frac{(\wt{x}_2-x_1+\lambda)(\wh{x}_2-x_2+\lambda)(\wt{x}_2-\wt{x}_1-\lambda)(\wt{x}_2-\wt{x}_2-\lambda)}{(\wt{x}_2-\wt{x}_1)(\wt{x}_2-x_1)(\wt{x}_2-x_2)}\;. 
\end{eqnarray}
Taking $\ln$, we get
\begin{eqnarray}
\ln |h_1|&=&\frac{1}{2}\left[ \ln | \lambda | + \ln | x_1-x_2+\lambda | + \ln |x_1-\wt{x}_1-\lambda| \right.\nn\\
&&\left.+\ln | x_1-\wt{x}_2-\lambda | - \ln | x_1-x_2 | -\ln | x_1-\wt{x}_1 | - \ln | x_1-\wt{x}_2 | \right],\;\;\; \\
\ln |\wt{h}_1|&=&\frac{1}{2}\left[ \ln | \lambda | + \ln | \wt{x}_1-x_1+\lambda | + \ln |\wt{x}_1-x_2+\lambda| \right.\nn\\
&&\left.-\ln | \wt{x}_1-\wt{x}_2-\lambda | - \ln | \wt{x}_1-\wt{x}_2 | -\ln | \wt{x}_1-x_1 | - \ln | \wt{x}_1-x_2 | \right],\;\;\; \\
\ln |h_2|&=&\frac{1}{2}\left[ \ln | \lambda | + \ln | x_2-x_1+\lambda | + \ln |x_2-\wt{x}_1-\lambda | \right.\nn\\
&&\left.+\ln | x_2-\wt{x}_2-\lambda |-\ln | x_2-x_1 | -\ln | x_2-\wt{x}_1 | - \ln | x_2-\wt{x}_2 | \right],\;\;\; \\
\ln |\wt{h}_2|&=&\frac{1}{2}\left[ \ln | \lambda | + \ln | \wt{x}_2-x_1+\lambda | + \ln |\wt{x}_2-x_2+\lambda| \right.\nn\\
&&\left.+\ln | \wt{x}_2-\wt{x}_1-\lambda | - \ln | \wt{x}_2-\wt{x}_1 | -\ln | \wt{x}_2-x_1 | - \ln | \wt{x}_2-x_2 | \right]\;.\;\; \;
\end{eqnarray}
Hence, 
\begin{eqnarray}
\det \boldsymbol M_{RS} &=& \ln | h_1 | + \ln | \wt{h}_1 | + \ln | h_2|+ \ln | \wt{h}_2| +\ln | \wt{x}_2-\wt{x}_1| \;\nn\\
&& + \ln | x_1-x_2 | - \sum\limits_{i,j = 1}^2 \ln | \wt{x}_i - x_j + \lambda | \; \nn\\
&&= 2 \ln | \lambda | - \sum\limits_{i,j = 1}^2 \ln | \wt{x}_i - x_j + \lambda | \; \nn\\
&&= \sum\limits_{i,j = 1}^2 \ln | x_i - x_j + \lambda | - \sum\limits_{i,j = 1}^2 \ln | x_i - \wt{x}_j |. \;
\end{eqnarray}
Obviously, for the $N$ particles or $N\times N$ matrix, we have
\begin{equation}
 \det \boldsymbol M_{RS} =  \sum\limits_{i,j = 1}^N \ln | x_i - x_j + \lambda | - \sum\limits_{i,j = 1}^N \ln | x_i - \wt{x}_j |  \;,
\end{equation}
which is indeed the discrete-time Lagrangian for the one-parameter CM system.
%%%%%%%%%%%%%%%%%%%%%%%%%%%%%%%%%%%%%%%%%%%%%%%%%%%%%%%%%%%%%%%
%%%%%%%%%%%%%%%%%%%%%%%%%%%%%%%%%%%%%%%%%%%
\begin{acknowledgements}

Umpon Jairuk would like to thank the Rajamangala University of Technology Thanyaburi(RMUTT) for financial support under the Personnel Development Fund in 2023.
\end{acknowledgements}
%%%%%%%%%%%%%%%%%%%%%%%%%%%%%%%%%%%%
%%%%%%%%%%%%%%%%%%%%%%%%%%%%%%%%%%%%%%%
%%%%%%%%%%%%%%%%%%%%%%%%%%%%%%%%%%%%

\end{document}